\documentclass[11pt]{article}
\usepackage{graphicx,citesort,ils2008}
\newcommand{\be}{\begin{equation}}
\newcommand{\ee}{\end{equation}}
\global\arraycolsep=2pt
\begin{document}

\begin{center}

{\LARGE\textbf{Coulomb Resummation and Monopole Masses}}

\vskip7.5mm

{\large K. A. Milton}

\vskip5mm

\textit{Oklahoma Center for High Energy Physics
and Homer L. Dodge Department of Physics and Astronomy\\
University of Oklahoma, Norman, OK 73019, USA}

\vskip10mm

\textbf{Abstract}

\vskip2.5mm

\parbox[t]{110mm}{{\small
The relativistic Coulomb resummation factor suggested by I.L. Solovtsov is
used to reanalyze the mass limits obtained for magnetic monopoles which might
have been produced at the Fermilab Tevatron.  
The limits given by the Oklahoma experiment (Fermilab E882) 
are pushed close to the unitary
bounds, so that the lower limits on monopole masses are increased from around 
250 GeV to about 400 GeV.
}}

\end{center}

\vskip5mm

\section{Threshold resummation}
\label{Sect1}
In describing a charged particle-antiparticle system near threshold, it is
well known from QED that the so-called
Coulomb resummation factor plays an important role. This resummation,
performed on the basis of the nonrelativistic
Schr\"odinger equation with the Coulomb potential $V(r)=-\alpha/r$,
leads to the Som\-mer\-feld-\-Sakha\-rov
$S$-factor \cite{ss}. In the threshold region one cannot
truncate the perturbative series and the
$S$-factor should be taken into account in its entirety. The $S$-factor
appears in the parametrization of the
imaginary part of the quark current correlator, which can be approximated
by the Bethe-Salpeter amplitude of the two
charged particles, $\chi_{\rm{BS}}(x=0)$ \cite{barbieri}.
The nonrelativistic replacement of this amplitude by
the wave function, which obeys the Schr\"odinger equation with the Coulomb
potential, leads to the appearance of the
resummation factor in the parametrization of the 
normalized $e^+e^-$ to hadrons cross section ratio $R(s)$.

For a systematic relativistic analysis of quark-antiquark systems,
it is essential from the very beginning to have a
relativistic generalization of the $S$-factor. A new form for this
relativistic factor in the case of QCD was
proposed by Solovtsov in 2001 \cite{ms01}.
\begin{equation}\label{S-factor-relativistic}
S(\chi)=\frac{X(\chi)}{1-\exp\left[-X(\chi)\right]}\, ,
\quad\quad X(\chi)=\frac{\pi\,\alpha}{\sinh\chi}\, ,
\end{equation}
where $\chi$ is the rapidity which is related to $s$ by
$2m\cosh\chi=\sqrt{s}$, $\alpha\to 4\alpha_s/3$ in QCD. The
function $X(\chi)$ can be expressed in terms of $v=\sqrt{1-4m^2/s}$:
$X(\chi)=\pi\alpha\sqrt{1-v^2}/v$. The
relativistic resummation factor (\ref{S-factor-relativistic}) reproduces
both the expected nonrelativistic and
ultrarelativistic limits and corresponds to a QCD-like Coulomb potential.
Here we consider the vector channel for
which a threshold resummation $S$-factor for the s-wave states is used. For
the axial-vector channel the $P$-factor
is required. The corresponding relativistic factor has been found recently
\cite{ssc05}.

To incorporate the quark mass effects one usually uses the approximate
expression above the quark-antiquark threshold \cite{pqw}
\begin{equation}\label{R-appr1}
{\cal{R}}(s)=T(v)\,\left[1+g(v)r(s)\right]\,,
\end{equation}
where
\begin{eqnarray}\label{vTg}
T(v)=v\frac{3-v^2}{2}\, ,\quad
g(v)=\frac{4\pi}{3}\left[\frac{\pi}{2v}-\frac{3+v}{4}
\left(\frac{\pi}{2}-\frac{3}{4\pi} \right) \right]\,  .
\end{eqnarray}
The function $g(v)$ is taken in the Schwinger approximation \cite{ss}.

\begin{figure}
\centerline{
\includegraphics[height=6cm]
{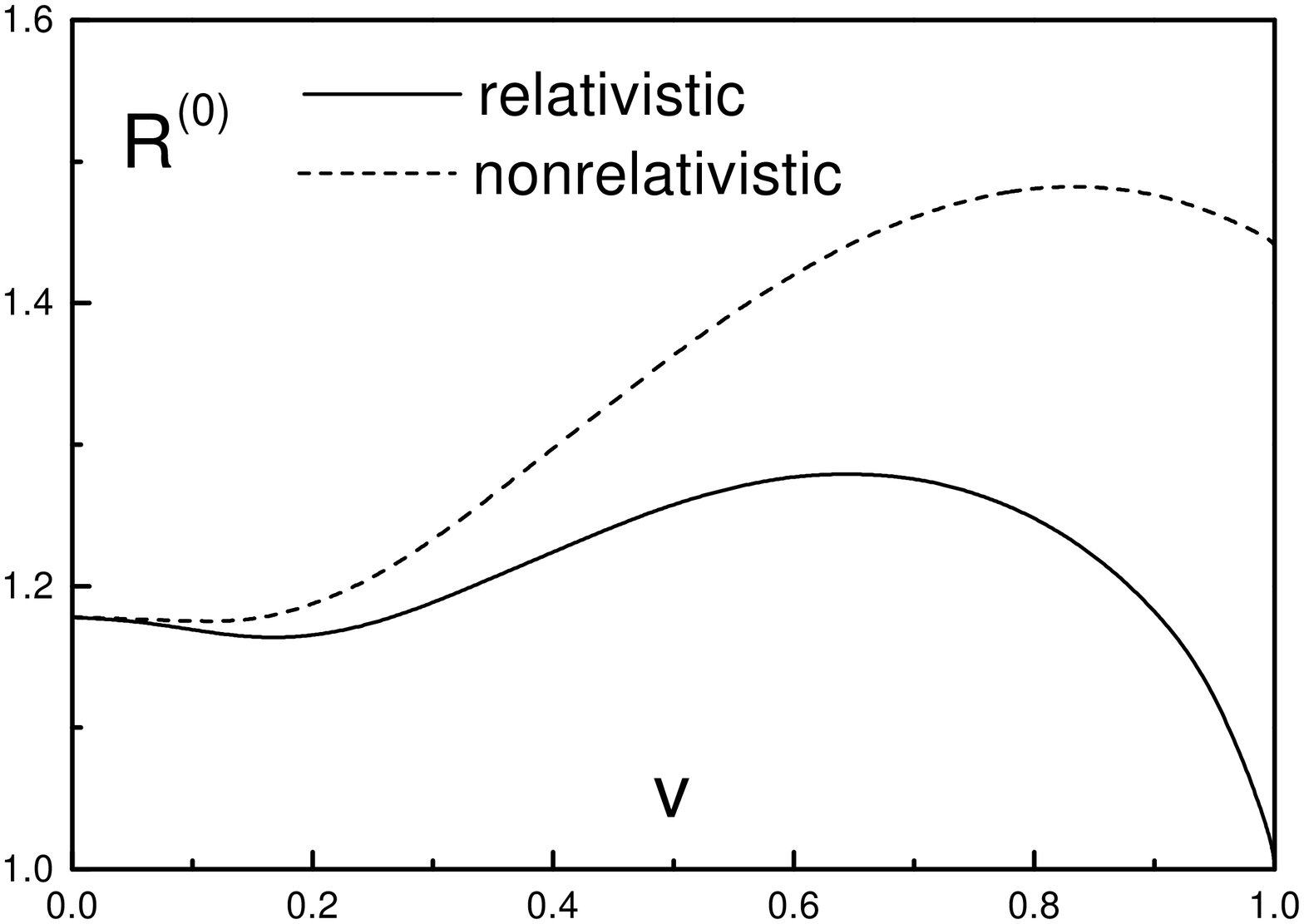}}
\caption{\sl Behavior of $R_0(s)$ with relativistic and nonrelativistic
$S$-factors.}
\label{fig2}
\end{figure}

One cannot directly use the perturbative expression for $r(s)$ in
Eq.~(\ref{R-appr1}), which contains unphysical
singularities, to calculate, for example, the Adler $D$-function.
Instead, one can use the analytic perturbation theory (APT) 
representation for $r(s)$. 
The explicit three-loop form for $r_{\rm APT}(s)$ can be found 
in Ref.~\cite{MSS-Adler-funct:01}.
Besides this replacement,
one has to modify the expression (\ref{R-appr1}) in such a way as to take into
account summation of an arbitrary number
of threshold singularities. Including the threshold resummation factor
(\ref{S-factor-relativistic}) leads to
the following modification of the expression (\ref{R-appr1})
for a particular quark flavor $f$ \cite{MSS-Adler-funct:01,sss}
\begin{eqnarray}
\label{R-R_0R_1}
{\cal{R}}_f(s)&=&\left[R_{0,f}(s)+R_{1,f}(s)\right]\Theta (s-4m_f^2), \\
R_0(s)&=&T(v)\,S(\chi), \qquad R_1(s)=T(v)\left[\,r_{\rm{APT}}(s)\,g(v)
-\frac{1}{2}X(\chi)\,\right]. \nonumber
\end{eqnarray}
%The function $r_{\rm{APT}}(s)$ is taken within the analytic approach
%as in~\cite{MSS-Adler-funct:01}.
The usage of the resummation factor (\ref{S-factor-relativistic})
reflects the assumption that the coupling is taken in the $V$
renormalization scheme. To avoid double counting, the function $R_1$
contains the subtraction of $X(\chi)$. The
potential term corresponding to the $R_0$ function gives the principal
contribution to ${\cal{R}}(s)$, as shown in Fig.~\ref{fig2}, the correction
$R_1$ amounting to less than twenty percent over the whole energy interval.
For a recent account of some of the successes of APT including Coulomb
resummation see Ref.~\cite{Milton:2005hp}.
%%%%%  FIGURE-2   ===  %%%%%%%%%%%%%%%%

\section{Dirac Magnetic Monopoles}
%\subsection{Interaction between electric and magnetic charge}
The relativistic interaction between an electric and a magnetic current
is \cite{miltonrev}
\begin{equation}
 W(j,{}^*j)=\int (d x) (d x') (d x'')
{}^*j^\mu(x)\epsilon_{\mu\nu\sigma\tau}
\partial^{\nu} f^{\sigma}\left(x-x^{\prime}\right)
D_{+}\left(x^{\prime}-x^{\prime\prime}\right)j^{\tau}
\left(x^{\prime\prime}\right).\label{interact}
\end{equation}
Here the electric and magnetic currents are
\be
j_\mu=e\bar\psi\gamma_\mu\psi\qquad \mbox{and}
\qquad{}^*j_\mu=g\bar\chi\gamma_\mu\chi,
\ee
for example, for spin-1/2 particles.
The photon propagator is denoted by $D_+(x-x')$ and $f_\mu(x)$ is
the Dirac string function which satisfies the differential equation
\begin{eqnarray}
\partial_\mu f^{\mu}(x)=4\pi\delta(x).\label{dfstng}
\end{eqnarray}
A formal solution of this equation is given by
\begin{eqnarray}
f^{\mu}(x)=4\pi n^\mu\left(n\cdot\partial\right)^{-1}\delta(x),\label{stngsl}
\end{eqnarray}
where $n^\mu$ is an arbitrary constant vector.

%\subsection{Charge quantization condition}
Dirac showed in 1931 \cite{dirac}
 that quantum mechanics was consistent with the existence
of magnetic monopoles provided the quantization condition holds,
\be
eg=m'\hbar c,\label{quant}
\ee
where $m'$ is an integer or an integer plus 1/2,
which explains the quantization of electric charge.
This was generalized by Schwinger to dyons, particles
carrying both electric charge $e_a$ and magnetic charge $g_a$ \cite{schwinger}:
\be
e_1g_2-e_2g_1=-m'\hbar c.\label{dyon}
\ee
(Schwinger sometimes argued 
that $m'$ was an integer, or perhaps an even integer.)
For details on the derivation of these quantization conditions,
see, for example, Ref.~\cite{miltonrev}.  In the following we write
$m'=n/2.$

\section{OU Monopole Experiment: Fermilab E882}
We now refer to the experiment conducted at the University of Oklahoma
from 1997--2004 \cite{kalbfleisch}, searching for low-mass monopoles which
might have been produced at the Tevatron and captured in the old CDF and D0
detectors.

The best prior experimental limit on the direct accelerator production
of magnetic monopoles is that
of Bertani et al.\ in 1990 \cite{bertani}
\be
\sigma\le2\times 10^{-34}\mbox{cm}^2 \quad\mbox{for a monopole mass}\quad
M\le850\,\mbox{GeV}.
\ee
(Such limits are complementary to searches for cosmic ``intermediate mass''
magnetic monopoles, with masses between $10^5$ and $10^{12}$ GeV, such
as have been recently reported in Ref.~\cite{Balestra:2008ps}.)
We are able to set much better limits than Bertani et al.\
because the integrated luminosity is $10^4$ times that of the previous 1990
experiment:
\be
\int {\cal L}=172\pm8\,\mbox{pb}^{-1} \quad(\mbox{D0}).
\ee
The fundamental mechanism is supposed to be a Drell-Yan process,
\be
p+\bar p\to q+\bar q+ X\to M+\bar M+X,
\ee
where the cross section is given by
\begin{eqnarray}\frac{d\sigma}{d \mathcal{M}}
=(68.5n)^2\beta^3\frac{8\pi\alpha^2}{9s}
\int \frac{d x_1}{x_1}\sum_iQ_i^2q_i(x_1)\bar q_i
\left(\frac{\mathcal{M}^2}{sx_1}\right).\label{dysigma}
\end{eqnarray}
Here $\mathcal{M}$
is the invariant mass of the monopole-antimonopole pair, and we have
included a factor of $\beta^3$
to reflect both phase space and  the velocity
suppression of the magnetic coupling, as roughly implied by
Eq.~(\ref{interact}).

\begin{figure}
\centering
\includegraphics[height=7cm]{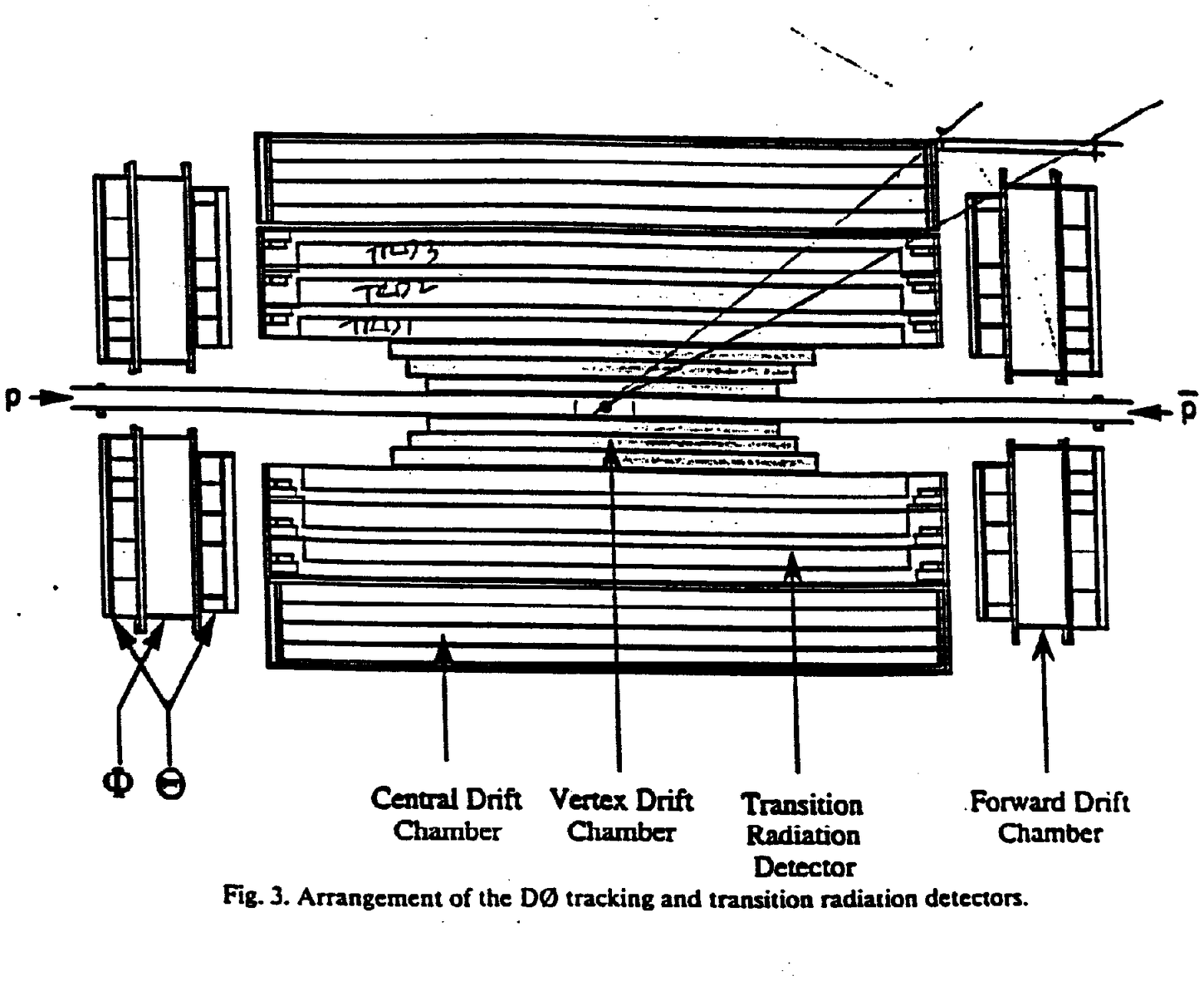}
\caption{\label{D0Detector} Arrangement of the D0 tracking and
transition radiation detectors.}
\end{figure}

Any monopole produced at the Tevatron is trapped in the detector elements with
100\% probability due to interaction with the magnetic moments of the
nuclei, based on the theory described in my review \cite{miltonrev}.
 The experiment consists of running samples obtained from the old
D0 and CDF detectors through a superconducting induction detector.
Figure \ref{D0Detector} is a sketch of the D0 detector.

We use energy loss formula of Ahlen \cite{ahlen} 
to describe the interaction of the monopoles with the detector elements.

\begin{figure}
\centering
\includegraphics[height=7cm]{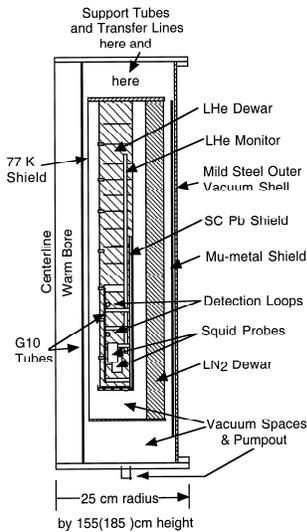}
\caption{\label{oudetector}
Sketch of the OU induction detector. Shown is a vertical cross
section; it should be imagined as rotated about the vertical axis labelled
``centerline.''}
\end{figure}

%\subsection{Sketch of OU induction detector}
Figure 
\ref{oudetector} is a diagram of the OU magnetic monopole induction detector.
It is a cylindrical detector, with a warm bore of diameter 10 cm, surrounded
by a cylindrical liquid N$_2$ dewar, which insulated a liquid He dewar.
The superconducting loop detectors were within the latter, concentric
with the warm bore.  Any current established in the loops was detected by a
SQUID.  The entire system was mechanically isolated from the
building, and magnetically isolated by $\mu$ metal and superconducting lead
shields.  The magnetic field within the bore was reduced with the help of
Helmholtz coils to about 1\% of the earth's field.
Samples were pulled vertically through the warm bore with a
computer-controlled stepper motor.  Each traversal took about 50 s; every
sample run consisted of some 20 up and down traversals.  Most samples
were run more than once, and more than 660 samples of Be, Pb, and Al from
both the old CDF and D0 detectors were analyzed over a period of 7 years.

\begin{figure}
\centering
\includegraphics[height=12cm]{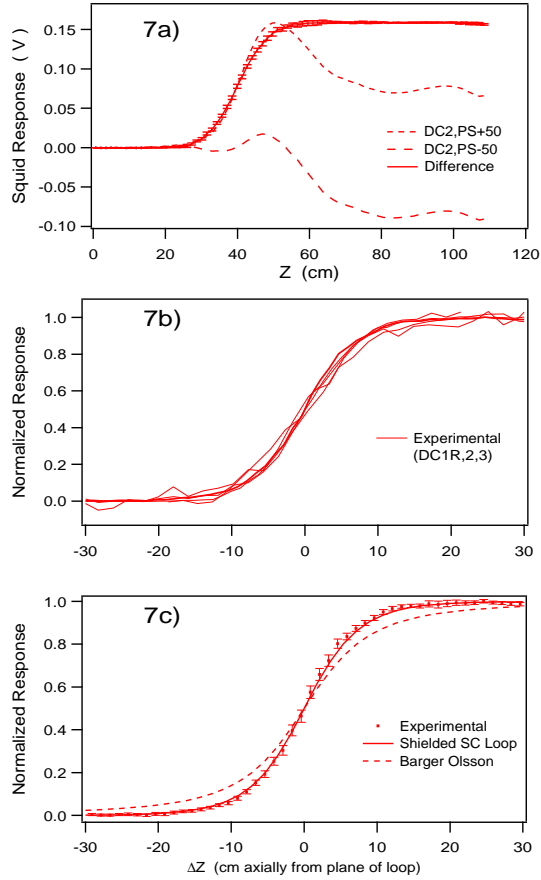}
\caption{\label{caldata}
Typical step plots: D0 aluminum, CDF lead, and CDF aluminum.
The experimental data was collected from pseudopole
simulations; the steps shown are for the difference between the results with
reversed polarizations of the pseudopole.  Data agrees well with the
SC theory,
which incorporates the effect of the shielded superconducting loops.
The theory without the shield, given by Barger and Ollson \cite{bo}
is also shown.}
\end{figure}

%\subsection{Signal}
A monopole passing through the superconducting loop would produce a
step in the current
\begin{equation}
LI=\frac{4\pi g}c-\frac{\Delta\Phi}c
=\frac{4\pi g}{c}\left(1-\frac{r^2}{a^2}\right).
\label{na17}
\end{equation}
where $L$ is the inductance of the loop, $r$ is the radius of the loop,
and $a$ is the radius of the superconducting cylinder.  The detector
was calibrated with a pseudopole, a long solenoid, and the resulting steps
in the output of the SQUID are seen in Fig.~\ref{caldata} to agree with theory. 

\begin{figure}
\centering
\includegraphics[height=6cm]{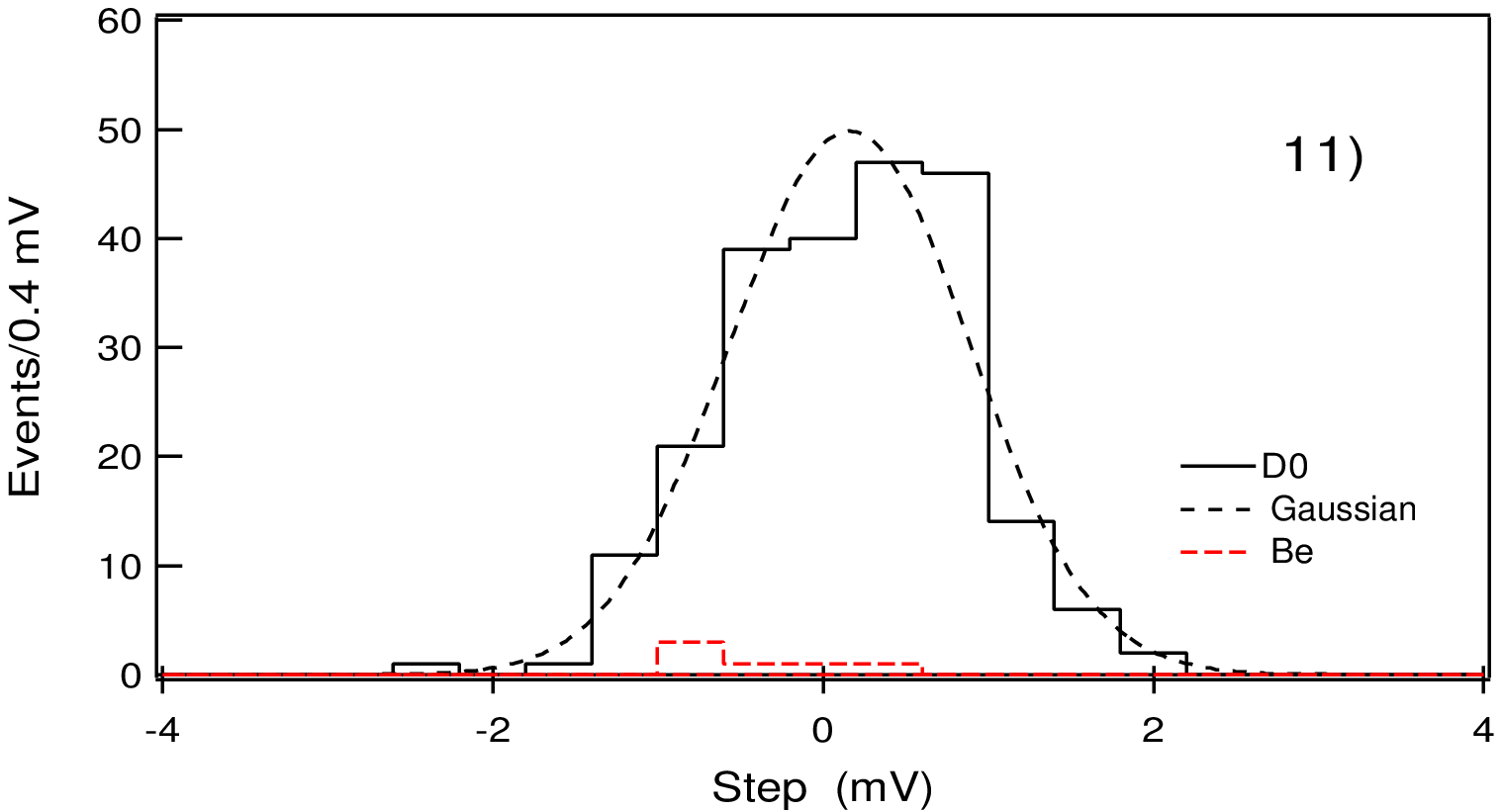}
\caption{\label{hist1}Steps from D0 samples.
A Dirac pole would appear as a step at 2.46 mV.}
\end{figure}

%\subsection{Histogram of steps}
Figure \ref{hist1} shows the histogram of steps from data collected from
D0 samples.  Similar histograms were obtained from the CDF data.

\begin{figure}
\centering
\includegraphics[height=12cm]{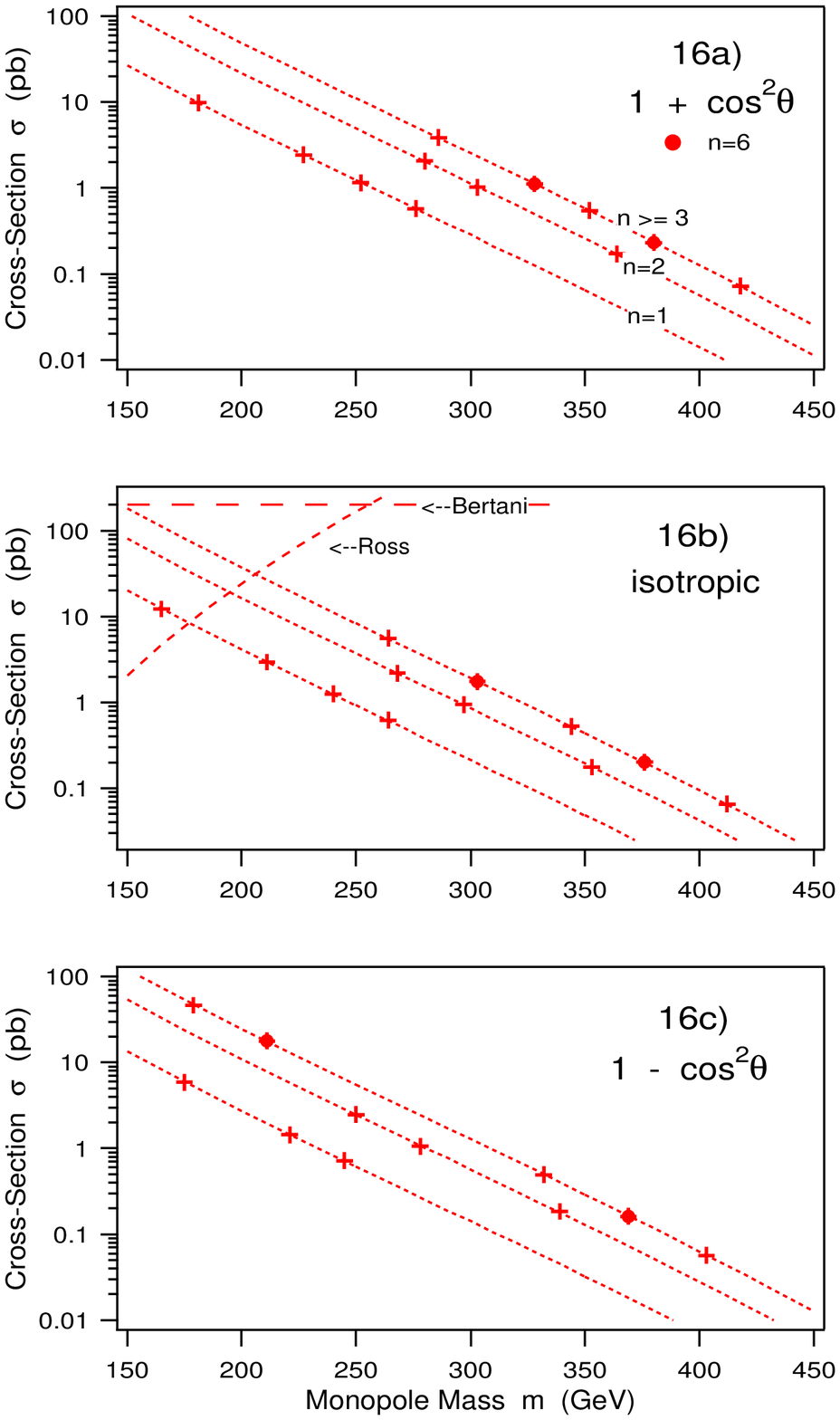}
\caption{\label{figmasslimits} Cross section vs.~mass limits.
The three graphs show three different assumptions about the angular
distribution, since even if we knew the spin of the monopole, we cannot
at present predict the differential cross section. Shown in the second
figure are the Bertani (1990) \cite{bertani} and lunar (1973) \cite{ross} 
limits.}
\end{figure}

%\subsection{Cross section versus mass limits}
From this, we can obtain limits on cross sections for the production of
monopole--antimonopole pairs, and then, model dependent limits on monopole
masses, as shown in Fig.~\ref{figmasslimits}.

%\subsection{Cross section and mass limits}

\begin{table}
\caption{\label{tablelimites}
Alternative interpretations for different production angular
distributions of the monopoles, comparing $1$ and $1\pm\cos^2\theta$
(90\% CL).
Here the upper cross section limits $\sigma_a$
corresponds to the distribution $1+a\cos^2\theta$, and similarly for the
lower mass limits $\mu$ (all at 90\% confidence level).}
\begin{tabular}{@{}llllllll}
\hline
Set&$2m'$&$\sigma_{+1}$&$\mu_{+1}$&
$\sigma_{0}$&$\mu_{0}$&
$\sigma_{-1}$&$\mu_{-1}$\\
&&(pb)&(GeV/$c^2$)&(pb)&(GeV/$c^2$)&(pb)&(GeV/$c^2$)\\
\hline
1 Al&1&1.2&250&1.2&240&1.4&220\\
1 Al RM&1&0.6&275&0.6&265&0.7&245\\
2 Pb&1&9.9&180&12&165&23&135\\
2 Pb RM&1&2.4&225&2.9&210&5.9&175\\
1 Al&2&2.1&280&2.2&270&2.5&250\\
2 Pb&2&1.0&305&0.9&295&1.1&280\\
3 Al &2&0.2&365&0.2&355&0.2&340\\
1 Be&3&3.9&285&5.6&265&47&180\\
2 Pb&3&0.5&350&0.5&345&0.5&330\\
3 Al&3&0.07&420&0.07&410&0.06&405\\
1 Be&6&1.1&330&1.7&305&18&210\\
3 Al&6&0.2&380&0.2&375&0.2&370\\
\hline
\end{tabular}
\end{table}

%\subsection{Summary of 2004 results}
Table \ref{tablelimites} shows the limits we obtained for different
sample sets, and different charges $m'$, for various assumed production
distributions.
Our best mass limits are (assuming isotropic distribution)
\begin{itemize}
\item $m'=\frac{1}2$:\quad $\mu>265$ GeV
\item $m'=1$:\quad $\mu>355$ GeV
\item $m'=\frac{3}2$:\quad $\mu>410$ GeV 
\item $m'=3$:\quad $\mu>375$ GeV.
\end{itemize}

\section{Reanalysis of Monopole Mass Limits Using Coulomb Resummation}

We will now use the Solovtsov Coulomb threshold correction (\ref{R-R_0R_1}) 
in the form
\be
R=T(v)[S(\chi)-\frac12X(\chi)],\label{rm}
\ee
with $T(v)$ given in (\ref{vTg})
%\be T(v)=v\frac{3-v^2}2,\ee
and $S(\chi)$ given in (\ref{S-factor-relativistic}), or
\be X(\chi)=\pi\alpha\frac{\sqrt{1-v^2}}v,\quad S(\chi)=\frac{X(\chi)}{1
-e^{-X(\chi)}}.
\ee
We have simply neglected the perturbative term, as uncalculable.

\begin{figure}
\centering
\includegraphics[height=6cm]{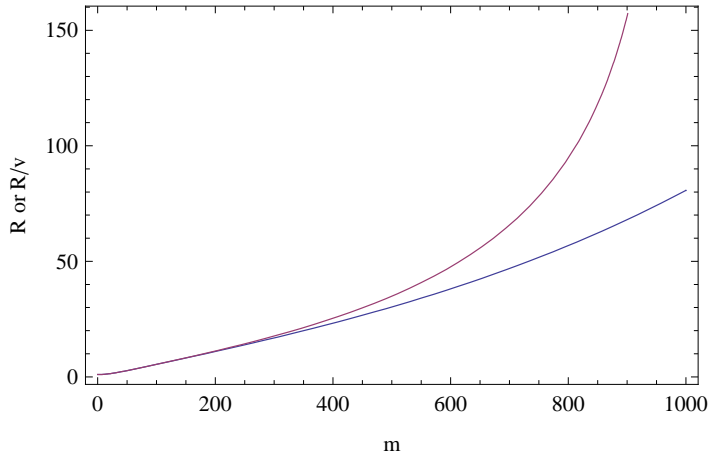}

\caption{\label{fig7}Enhancement factor $R$ for the theoretical monopole 
cross section
due to Coulomb resummation, as a function of monopole mass $\mu=\frac{
\mathcal{M}}2 m$, where $\mathcal{M}$ is the invariant mass of the
monopole-antimonopole pair in TeV, and $\mu$ and $m$ are in GeV.
The bottom curve shows $R$ given in Eq.~(\ref{rm}), while in the top
curve this
factor is divided by the monopole velocity, since that is already
included in the analysis, see Eq.~(\ref{dysigma}).}
\end{figure}

This is a small correction in QED, but here from Eq.~(\ref{quant})
\be
\alpha\to 137 \left(\frac{n}2\right)^2,\quad n=1,2,3,\dots.
\ee
Figure \ref{fig7} shows the substantial resulting increase in the cross
section.  This essentially pushes the cross section to the unitarity
limit,
\be
\sigma \le \frac{\pi(2J+1)}s\sim \frac{3\pi}s.
\ee
As a result, for all charge states, our limits become
\be
\mu>400 \,\mbox{GeV}.
\ee

\section{Conclusions}
The relativistic Coulomb resummation factor plays an important
role in analysis of QCD experiments.
Because the coupling is strong, it also plays a significant role
in the theory of the production of magnetic monopole--anti-monopole pairs.
Of course, because of the strong coupling, and even more because of
the nonperturbative aspects of the Dirac string, there are potentially
other effects which are just as strong but uncalculable.
Our estimates of production rates were therefore extremely conservative,
and a realistic assessment of the situation suggests that the limits on
monopole masses from the Oklahoma experiment are at least as strong as
the published limit from the very different CDF experiment \cite{cdf}:
\be
\mu_{\rm CDF} \ge 360 \,\mbox{GeV},
\ee

\section*{Acknowledgements} This work was supported in part by a grant from
the US Department of Energy.  I dedicate this paper to the memory of my dear
friend and colleague, Igor Solovtsov, and also to the memory of George
Kalbfleisch.  Both left us far too early, and are sorely missed.


\begin{thebibliography}{99}

\bibitem{ss} A. Sommerfeld, {\it Atombau and Spektallinien}, vol.~2
(Vieweg, 1939); A. D. Sakharov, Zh.\ Eksp.\ Teor.\ Fiz.\ {\bf 18}, 631 (1948);
J. Schwinger, {\it Particles, Sources, and Fields}, vol.~2 (Addison-Wesley,
1973, Perseus, 1998).
\bibitem{barbieri} R. Barbieri, P. Christillin, and E. Remiddi, Phys.\
Rev.\ D {\bf 8}, 2266 (1973).
\bibitem{ms01} K. A. Milton and I. L. Solovtsov, Mod.\ Phys.\ Lett.\ A 
{\bf16}, 2213 (2001)
\bibitem{ssc05} I. L. Solovtsov, O. P. Solovtsova, and Yu. D. Chernichenko,
Phys.\ Part.\ Nucl.\ Lett.\ {\bf 2}, 199 (2005).
\bibitem{pqw} E. C. Poggio, H. R. Quinn, and S. Weinberg, Phys.\ Rev.\ D
{\bf 13}, 1958 (1976); T. Appelquist and H. D. Politzer, Phys.\ Rev.\ Lett.\
{\bf 34}, 43 (1975); Phys.\ Rev.\ D {\bf 12}, 1404 (1975).
\bibitem{MSS-Adler-funct:01} K. A. Milton, I. L. Solovtsov, and O. P.
Solovtsova, Phys.\ Rev.\ D {\bf 64}, 016005 (2001).
\bibitem{sss} A. N. Sissakian, I. L. Solovtsov, and O. P. Solovtsova,
JETP Lett.\ {\bf 73}, 166 (2001).

\bibitem{Milton:2005hp}
  K.~A.~Milton, I.~L.~Solovtsov and O.~P.~Solovtsova,
  %``An analytic method of describing R-related quantities in QCD,''
  Mod.\ Phys.\ Lett.\  A {\bf 21}, 1355 (2006)
  [arXiv:hep-ph/0512209].
  %%CITATION = MPLAE,A21,1355;%%
\bibitem{miltonrev} K. A. Milton, %``Theoretical
%and experimental status of magnetic monopoles,'' 
Rep.\ Prog.\ Phys.\
{\bf 69}, 1637 (2006).
\bibitem{dirac} P. A. M. Dirac, Proc.\ R. Soc. London A {\bf 133}, 60 (1931).
\bibitem{schwinger} J. Schwinger, Science {\bf 165}, 757 (1969).
\bibitem{kalbfleisch} Kalbfleisch et al., Phys.\ Rev.\ Lett.\ {\bf 85}, 5292 (2000);
Phys.\ Rev.\ D {\bf 69}, 052002 (2004).
\bibitem{bertani} M. Bertani et al., Europhys.\ Lett.\ {\bf 12}, 613 (1990).

\bibitem{Balestra:2008ps}
  S.~Balestra {\it et al.},
  %``Magnetic Monopole Search at high altitude with the SLIM experiment,''
  arXiv:0801.4913 [hep-ex].
  %%CITATION = ARXIV:0801.4913;%%
\bibitem{ahlen} S. P. Ahlen and K. Kinoshita, Phys.\ Rev. D {\bf 26}, 2347
(1982); S. P. Ahlen, in {\it Magnetic Monopoles}, ed.~R. A. Carrigan and W. P.
Trower (New York, Plenum, 1982), p.~ 259.
\bibitem{bo} V. Barger and M. G. Ollson, {\it Classical Electricity and
Magnetism} (Boston: Allyn and Bacon, 1967).
\bibitem{ross} R. R. Ross, P. H. Eberhard, L. W. Alvarez, and R. D. Watt,
Phys.\ Rev.\ D {\bf 8}, 698 (1973).
\bibitem{cdf} A.~Abulencia {\it et al.}  [CDF Collaboration],
  %``Direct search for Dirac magnetic monopoles in $p\bar{p}$ collisions at
  %$\sqrt{s} = 1.96$ TeV,''
Phys.\ Rev.\ Lett.\  {\bf 96}, 201801 (2006)  [arXiv:hep-ex/0509015].
  %%CITATION = PRLTA,96,201801;%%


\end{thebibliography}
\end{document}